\documentclass[aps,prb,reprint,amsmath,amssymb,superscriptaddress,floatfix]{revtex4-2}
\usepackage{graphicx,xcolor,hyperref}
\usepackage[ignoreunlbld,norefs,nocites]{refcheck}
\begin{document}

\title{Quantum Monte Carlo study of the quasiparticle effective mass of the two-dimensional uniform electron liquid}
\author{S.\ Azadi}
\email{sam.azadi@manchester.ac.uk}
\affiliation{Department of Physics and Astronomy, University of Manchester, Oxford Road, Manchester M13 9PL, United Kingdom}
\author{N.\ D.\ Drummond}
\affiliation{Department of Physics, Lancaster University, Lancaster LA1 4YB, United Kingdom}
\author{A.\ Principi}
\affiliation{Department of Physics and Astronomy, University of Manchester, Oxford Road, Manchester M13 9PL, United Kingdom}
\author{R.\ V.\ Belosludov}
\affiliation{Institute for Materials Research, Tohoku University, Sendai 980-08577, Japan}
\author{M.\ S.\ Bahramy}
\affiliation{Department of Physics and Astronomy, University of Manchester, Oxford Road, Manchester M13 9PL, United Kingdom}
\date{\today}

\begin{abstract}
The real-space variation quantum Monte Carlo (VMC) and diffusion quantum Monte Carlo (DMC) are used to calculate the quasiparticle energy bands and the quasiparticle effective mass of the paramagnetic and ferromagnetic two-dimensional uniform electron liquid (2D-UEL)\@. The many-body finite-size errors are minimized by performing simulations for three system sizes with the number of electrons $N=146$, 218, and 302 for paramagnetic and $N=151$ for ferromagnetic systems. We consider 2D-UEL to be within the metallic density range $1\leq r_s \leq 5$. The VMC and DMC results predict that the quasiparticle effective mass $m^*$ of the paramagnetic 2D-UEL at high density $r_s=1$ is very close to 1, suggesting that effective mass renormalization due to electron-electron interaction is negligible. We find that $m^*$ of the paramagnetic 2D-UEL obtained by the VMC and DMC methods increases by $r_s$ but with different slopes. Our VMC and DMC results for ferromagnetic 2D-UEL indicate that $m^*$ decreases rapidly by reducing the density due to the strong suppression of the electron-electron interaction. 
\end{abstract}

\maketitle

\section{Introduction}
The two-dimensional uniform electron liquid (2D-UEL) is one of the most fundamental models in physics because it captures the essential physics of interacting electrons in two dimensions, which is a regime where quantum mechanics, statistics, and Coulomb interactions all play equally critical roles \cite{Pines66,Loos16,Dreizler,Hohenberg64,Kohn65,Landau,Luttinger61,Senatore94,Dornheim16,Dornheim18,Vosko80,Perdew92,Bhattarai18, Holzmann20, Holzmann16,Neilprb09,Neil08,Tantar89,Azadi24,Kwon94,Kwon96,Asgari06,Asgari09,Holzman09,Neilprl09,Neilprb13,Calvera2025,Gino}. Although it is simple, it is nontrivial as strong interactions and quantum effects cannot be treated exactly except in special limits. This combination makes it a benchmark for developing, testing, and understanding many-body theories. In fact, it is a playground to test our understanding of quantum many-body physics.

In a 2D-UEL the strong electron-electron correlations modify the behavior of individual electrons, leading to the emergence of quasiparticles with renormalized properties such as energy dispersion, effective mass $m^*$, and lifetime. These properties can be understood within the framework of Fermi-liquid theory \cite{Landau}. There is much experimental evidence for the renormalization of electronic properties. For instance, the effective mass renormalization can be measured using Shubnikov-de Haas (SdH) oscillations, cyclotron resonance experiments, and angle-resolved photoemission spectroscopy (ARPES), which directly probes the renormalized dispersion relation \cite{Padmanabhan08,Gokmen09,Tan05,Smith72}. The $m^*$ governs how electrons respond to external forces and determines the group velocity $v_F = k_F/m^*$, where $k_F$ is the Fermi momentum. Hence, $m^*$ affects the current, conductivity, and collective modes \cite{Giuliani05}. 

In a simple noninteracting 2D-UEL, the Hamiltonian subjected to periodic boundary conditions (PBC) contains only one-body term $H_0=\sum_{i=1}^N {p_i^2}/{2m}$ where $p_i$ and $m$ are the momentum operator of the electron $i$, and bare electron mass, respectively. Within this simplistic description of a Fermi liquid, $H_0$ can be written as the product of single-particle wave functions, each of which satisfies the Schr\"{o}dinger equation
$({-\nabla^2}/{2m})\psi_{\mathbf{k}}(\mathbf{r})=\varepsilon_{\mathbf{k}}\psi_{\mathbf{k}}(\mathbf{r})$, where for simplicity we ignored electron spin and used atomic units $\hbar=1$. $\mathbf{k}$ and $\mathbf{r}$ represent the set of quantum numbers characterizing the one-particle quantum state and the electron vector position, respectively. The eigenfunction of $H_0$ with eigenvalue $E$ is a fully antisymmetric wave function under exchange of the coordinates of two particles with combinations $\Psi(\mathbf{r}_1,\dots,\mathbf{r}_N)=\frac{1}{\sqrt{N!}}\sum_P(-1)^P P\Pi_{i=1}^N \psi_{k_i}(\mathbf{r}_i)$. This wave function satisfies the Pauli exclusion principle, namely the occupation number of the single-particle state $\psi_{k_i}$ is 0 or 1. This can be seen by writing the wave function in determinant form, called a Slater determinant:
\begin{equation}\label{eq:slater}
    \Psi(\{\mathbf{r}\})=\frac{1}{\sqrt{N!}}\det 
    \begin{vmatrix}
    \psi_{k_1}(\mathbf{r}_1) & \psi_{k_1}(\mathbf{r}_2)& \dots & \psi_{k_1}(\mathbf{r}_N) \\
    \psi_{k_2}(\mathbf{r}_1) & \psi_{k_2}(\mathbf{r}_2)& \dots & \psi_{k_2}(\mathbf{r}_N) \\
    \vdots & \vdots & \cdots & \vdots \\
    \psi_{k_N}(\mathbf{r}_1) & \psi_{k_N}(\mathbf{r}_2)& \dots & \psi_{k_N}(\mathbf{r}_N)
    \end{vmatrix}
\end{equation}
The determinant is zero if two sets of quantum numbers are equal (i.e., two rows are the same) and changes sign under the exchange of the coordinates of two fermions (i.e., exchange of two columns). As imposed by PBC the single-particle states are plane waves $\psi_i=e^{i\mathbf{k}\cdot \mathbf{r}}/\sqrt{A}$ where $i\equiv k_x, k_y$, and $e^{ik_x(x+L)}=e^{ik_xx}, \space e^{ik_y(y+L)}=e^{ik_yy}$ where $L$ and $A$ are the side and surface of the simulation cell, respectively. Hence, the electron momentum $k$ is discretized and can only have the value of $k_\alpha=\frac{2\pi}{L}n_\alpha; \space \alpha=x,y$ where $n_\alpha$ is an integer number. The single particle and total energies of the system are given by $\varepsilon_{\mathbf{k}}=\frac{\mathbf{k}^2}{2m}$ and $E=\sum_{i=1}^N \varepsilon_{\mathbf{k}_i}$, respectively. 

In the non-interacting electron model of 2D-UEL, the ground state of the $N$-electron system is obtained by occupying the $N$ lowest energy one-particle state characterized by a set of quantum numbers $\mathbf{k}_1,\mathbf{k}_2,\dots,\mathbf{k}_N$. The highest occupied energy state defines the Fermi level $k_F$, and its energy defines the Fermi energy $\varepsilon_F=k_F^2/2m$. Hence, the ground state is created by filling a circle with radius $k_F$. The value of $k_F$is related to the density $\rho=N/A$ via $\rho=\frac{k_F^2}{2\pi}$. It can be easily obtained that the energy per particle for a noninteracting 2D-UEL is $E/N = k_F^2/4m$. Almost all properties of 2D-UEL can be described as a function of the density parameter $r_s$ and spin-polarization $\zeta$ which play a crucial role in the physics of the electron liquid and are defined as $r_s=(1/\rho\pi)^{1/2}a_B^{-1}$, where $\rho$ is the electron number density and $a_B=1$ a.u.\ is the Bohr radius, $\zeta=(N_\uparrow-N_\downarrow)/N$ in which $N_\uparrow$, $N_\downarrow$, and $N$ are the number of electrons with spin up and down and the total number of electrons in the system. It can be easily concluded that the Fermi wave vectors of the non-spin polarized $\zeta=0$ (paramagnetic) and fully spin polarized $\zeta=1$ (ferromagnetic) 2D-UEL are $k_{F}^{\zeta=0}=\sqrt{2}/r_s$ and $k_{F}^{\zeta=1}=2/r_s$, respectively. The ferromagnetic system has a larger $k_F$, since all electrons occupy a single spin band, increasing the required area in the $k$-space.

When Coulomb electron-electron interactions are included, the energy dispersion deviates from the free-electron form. The many-body effects modify the energy dispersion as: $E(k) = {\hbar^2 k^2}/{2m} + \Sigma(k, \omega)$ where $\Sigma(k, \omega)$ is the self-energy correction due to interactions including exchange and correlation causing the renormalization of the bare electron mass. In real materials, the renormalized mass of electrons, called effective mass $m^*$, differs from the bare mass due to electron-electron interactions, electron-phonon interactions, and plasmonic effects. Although scattering with lattice vibrations and coupling with collective excitations modify $m^*$, at very low temperature, electron-electron Coulomb interactions are the dominant scattering mechanism, leading to a finite lifetime $\tau_{\text{ee}}^{-1} \sim E^2$ indicating Fermi-liquid behavior. According to Fermi-Liquid theory \cite{Landau} quasiparticle excitations can be characterized by the renormalization constant $Z(k_F)$, which is related to the residue of Green's function at $k=k_F$ and $m^*$. In a simple picture, $1-Z(k_F)$ and $1-m^*/m$ both provide information about the amount of many-body effects in the Fermi liquid.  

A common way to estimate the renormalized mass is using the many-body perturbation theory to calculate the self-energy function $\Sigma(k, \omega)$, which was initially introduced by Hedin ($GW$ approximation) \cite{Hedin1965}. We summarize the formalism here to clarify the reason behind the recent contradiction between the $m^*$ results for 3D UEL \cite{Azadi21,Holzmann2023,Haule2022}. The standard staring point of $GW$ calculations is the Dyson equation for the Green's function:
\begin{equation}\label{eq:DysonGreen}
    G_\sigma(\mathbf{k},\omega)=[\omega - \varepsilon_{\mathbf{k}}^0 - \Sigma(\mathbf{k},\omega)]^{-1}
\end{equation}
where $\varepsilon_{\mathbf{k}}^0$ and $\Sigma(\mathbf{k},\omega)$ are non interacting single particle energy and the irreducible self-energy, respectively. The effective mass which is a characteristic feature of the quasiparticle excitation dispersion curve $\varepsilon_{\mathbf{k}}={\mathbf{k}^2}/{2m^*}$ can be written as \cite{Ichimaru1994}
\begin{equation} %\label{eq:effectmass}
    \frac{m^*}{m}=\frac{1-\left.\frac{\partial \Sigma(\mathbf{k},\omega)}{\partial \omega}\right|_{k=k_F}}{1+\left.\frac{m}{k}\frac{\partial \Sigma(\mathbf{k},\omega)}{\partial k}\right|_{k=k_F}}
\end{equation}
where $m$ is bare electron mass. Using the Dyson equation, self-energy is given as
\begin{equation}\label{eq:DysonSelf}
    \Sigma(\mathbf{k},\omega)=i\int \frac{d\mathbf{q}}{(2\pi)^3} \int W(\mathbf{q}, \omega^\prime)G_\sigma(\mathbf{k}-\mathbf{q}, \omega-\omega^\prime)\frac{d\omega^\prime}{2\pi}
\end{equation}
the many-body effects are included in $W$ function, which can be approximated by
\begin{equation}\label{eq:WFunct}
    W(q,\omega)=\frac{v_q}{\epsilon(q,\omega)}\Gamma(q,\omega)
\end{equation}
where $v_q=4\pi/{q^2}$ is the bare Coulomb interaction, $\epsilon(q,\omega)$ is the exact dielectric function, and $\Gamma(q,\omega)$ is the vertex correction \cite{Ichimaru1994}. Hedin used the random phase approximation (RPA) in his work by choosing $\Gamma=1$ and using the RPA dielectric response in Eq.~(\ref{eq:WFunct}). Therefore, $W$ was an effective RPA screened interaction
\begin{equation}\label{eq:RPA}
    W^{RPA}(q,\omega)=\frac{v_q}{\epsilon^{RPA}(q,\omega)}
\end{equation}
The self-energy $\Sigma(\mathbf{k},\omega)$ was calculated by substituting the $W$ function from Eq.~(\ref{eq:RPA}) into Eq.~(\ref{eq:DysonSelf}) and using the non-interacting Green's function $G_0$ in the right-side of Eq.~(\ref{eq:DysonSelf}). Within this $G_0W_0$ approximation, as was named in Hedin's work, $m^*/m$ of 3D-UEL increases within the metallic regime and becomes larger than 1 for $r_s>3$. A similar behavior for $m^*/m$ of 3D-UEL was observed in a recent work in which static self-energy $\Sigma(k,0)$ was obtained using variational Monte Carlo (VMC) \cite{Holzmann2023}, in agreement with variational diagrammatic Monte Carlo simulations \cite{Haule2022,Chen2019}. However, if Eqs~(\ref{eq:DysonGreen}) and (\ref{eq:DysonSelf}) are solved self-consistently, even within the RPA approximation, $m^*/m$ of 3D-UEL monotonically decreases with $r_s$ as reported in previous works \cite{Rietschel,Yasuhara,Nakano89,Nakano89II,Krakovsky1995}, as we also observed in our quantum Monte Carlo (QMC) simulations of the quasiparticle energy bands of 3D-UEL \cite{Azadi21}. Hence, in this work we use a methodology similar to our previous work in which the DMC method is used to calculate the energy dispersion curve and the $m^*$ of 3D-UEL \cite{Azadi21}. In this work both the VMC and DMC methods are employed to obtain the energy bands and $m^*$ of the paramagnetic and ferromagnetic 2D-UEL within the metallic density range $1\leq r_s \leq 5$. 

\section{Details of the QMC calculations}
We used the Slater-Jastrow (SJ) and SJ-backflow (SJB) trial wave functions. The SJ wave function has the form $\Psi(\mathbf{R}_\uparrow, \mathbf{R}_\downarrow)=e^{J(\mathbf{R}_\uparrow, \mathbf{R}_\downarrow)}D_\uparrow(\mathbf{R}_\uparrow)D_\downarrow(\mathbf{R}_\downarrow)$ where $D_\uparrow$ and $D_\downarrow$ are Slater determinants of up- and down-spin single-particle plane wave orbitals defined in Eq.~(\ref{eq:slater}). $\mathbf{R}_\uparrow$, $\mathbf{R}_\downarrow$, and $e^J$ are the vector coordinates of the up- and dow-spin electrons, and the Jastrow correlation factor, respectively. In the SJB wave function the electron coordinates in the Slater determinant were replaced by the coordinates obtained by backflow (BF) transformation represented by a plolynomial in the electron-electron distance \cite{Pablo06,Pablo12}. The Jastrow term consisted of polynomial and plane wave expansions in electron-electron separation \cite{Neil04,Neil08,Neilprb09}. The variational parameters in the trial wave function were optimized using variance minimization \cite{Umrigar88,Neil05} followed by linear least square energy minimization \cite{Umrigar} as implemented in the CASINO package \cite{casino}. The quality of our optimized wave function is such that the VMC and DMC results for $m^*$ are equal within the error bar. 

The VMC and DMC energy bands $\varepsilon(\mathbf{k})$ are calculated by evaluating the difference in QMC energy when an electron is added to or removed from a state with momentum $\mathbf{k}$. The single-particle energy for an occupied and unoccupied state at momentum $\mathbf{k}$ is defined as $\varepsilon_-(\mathbf{k})= E_0 - E_-(\mathbf{k})$ and $\varepsilon_+(\mathbf{k})= E_+(\mathbf{k})-E_0$, respectively, where $E_0$ is the ground state total energy, $E_-(\mathbf{k})$ ($E_+(\mathbf{k})$) is total energy of the system with an electron removed from (added to) the single particle orbital $e^{i\mathbf{k}\cdot \mathbf{r}}$ in the Slater determinant defined in Eq.~(\ref{eq:slater}). Electronic excitations near the Fermi level correspond to quasiparticle excitations. The electronic and quasiparticle bands agree vicinity of the Fermi surface and therefore have the same derivative at $k_F$. The effective mass is calculated using the derivative of the energy bands $m^*=k_F.({d\varepsilon}/{dk})_{k_F}^{-1}$. We minimized FS errors by performing calculations for different simulation cell sizes with up to 302 electrons. We determined the energy band at a series of $k$ values and carried out a least-square fit of a Pad\'{e} function $\varepsilon(k)=(a_0+a_1k+a_2k^2+a_3k^3)/(1+2a_3k)$ and a quartic function $\varepsilon(k)=a_0+a_2k^2+a_4k^4$ to the band values. We discussed how the final results for $m^*$ depend on the fitting function. 

The QMC simulations were performed using a finite simulation cell with hexagonal symmetry subject to PBC\@. Hence, the available momentum states $\mathbf{k}$ are located on the grid of reciprocal lattice points offset by the Bloch vector of the simulation cell $\mathbf{k}_s$, which we set zero in our calculations. We used $r_s^\prime=r_{s_0}\sqrt{N_0/N^\prime}$ where $N^\prime=N-1$ and $N^\prime=N+1$ to calculate $\varepsilon_-(\mathbf{k})$, and $\varepsilon_+(\mathbf{k})$, respectively. Hence, the simulation cell volume was fixed when an electron was removed from or added to the simulation cell. 

We studied paramagnetic and ferromagnetic 2D-UEL at densities $r_s=1,2,3,4,5$ and performed VMC and DMC calculations for simulation cells containing $N=146$, 218, and 302 electrons for the paramagnetic system and $N=151$ for the ferromagnetic system. We calculated the VMC and DMC quasiparticle energies at more than twenty momentum vectors within the range $0< k < 1.7k_F$. In the thermodynamic limit, the exact energy band is smooth and well behaved near the Fermi level. However, the Hartree-Fock (HF) band is pathological, as discussed in the next section. In a finite system size, the HF band oscillates at momentum vectors near the $k_F$, and although QMC retrieves a large fraction of the correlation energy, it does not fully eliminate the pathological behavior of HF theory. Therefore, it is important to consider excitations away from the Fermi surface to obtain the gradient of the quasiparticle energy band at $k_F$.

\section{Results and discussion}
In the following sections, we systematically analyze different factors that can affect the QP energy bands obtained by VMC and DMC\@. The single particle energy of occupied and unoccupied states is obtained using subtraction of two energies, and hence any source of error or improvement in calculations may cancel out due to subtraction. In the following, we discuss when this can or cannot be true. 

\subsection{Hartree-Fock energy band} %\label{sec:HF}
\begin{figure}[!htb]
    \centering
        \includegraphics[width=1.0\linewidth]{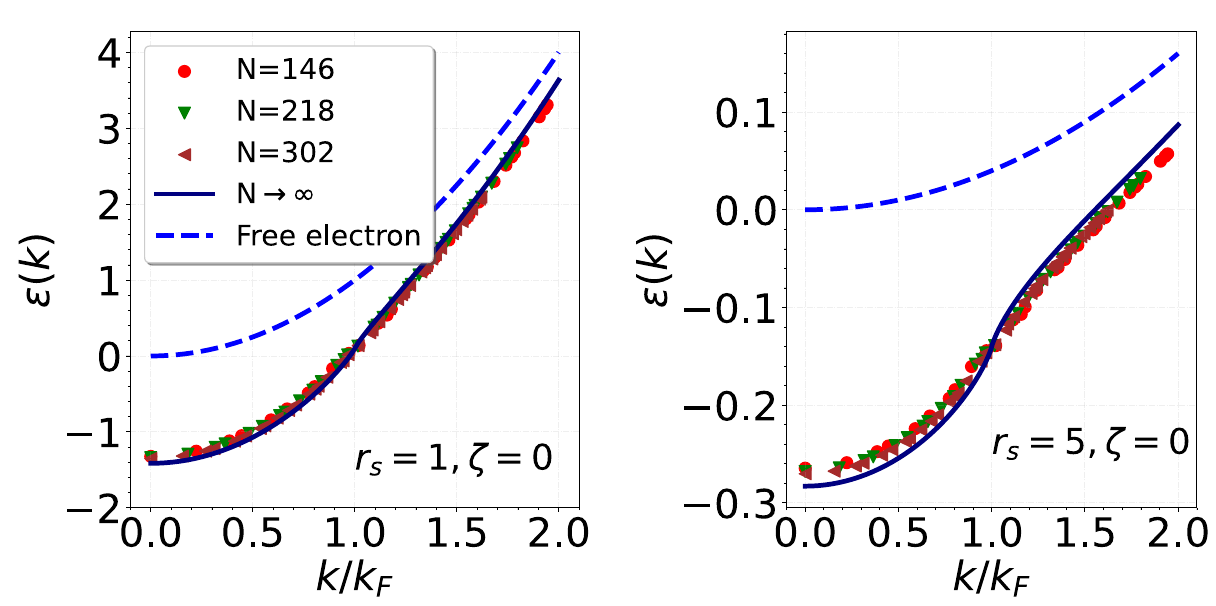}
    \caption{HF energy bands in (a.u.)\ for $N$-electron ($N=146$, 218, and 302) paramagnetic ($\zeta=0$) 2D-UEL with density parameter $r_s=1, 5$. HF energy band in the infinite system size limit (N$\rightarrow \infty$) defined in Eq.~(\ref{eq:HF}) is also plotted. Legend for plot $r_s=5$ is the same as $r_s=1$.}
    \label{fig:HFband}
\end{figure}

According to HF theory, the energy of a single electron with spin $\sigma$  is given by $\varepsilon_{\sigma}(\mathbf{k}) 
=\varepsilon^0(\mathbf{k})+\varepsilon^{(x)}_{\sigma}(\mathbf{k})$, where the first and second terms are the free-electron energy and the exchange contribution, respectively. Within HF theory, the two components of spin are not mixed, and therefore the ground state of HF is given by the product of two independent full Fermi circles with radii $k_{F_\uparrow}$ and $k_{F_\downarrow}$. The exchange term for 2D-UEL can be written as \cite{Giuliani05}
\begin{equation}
    \varepsilon^{(x)}_{\sigma}(\mathbf{k})=-\int \frac{1}{|\mathbf{k}-\mathbf{q}|}\frac{d^2q}{2\pi}=-\frac{2k_{F,\sigma}}{\pi}f(k/k_{F,\sigma})
\end{equation}
where the function $f$ is given by
\begin{equation}\label{eq:HF}
    f(x)=
    \begin{cases}
    E(x) & x\leq 1 , \\
    x[E(\frac{1}{x})-(1-\frac{1}{x^2})K(\frac{1}{x})] & x\geq 1.
    \end{cases}
\end{equation}
where $K(x)$ and $E(x)$ are the complete elliptic integrals respectively of the first and second kind. The HF energy band of 2D-UEL at density parameters $r_s=1, 5$ is shown in Fig.\ref{fig:HFband}. It can be observed that the HF bands of finite systems and the system in the thermodynamic limit (N$\rightarrow \infty$) exhibit pathological behavior close to $k=k_F$. Hence, the derivatives of bands near the Fermi surface become extremely large and fluctuate. This pathological behavior at the Fermi surface affects the DMC energy bands because not all the correlation energy is retrieved by the DMC even when a BF is used. Furthermore, because of the logarithmic divergence of ${d\epsilon^{(x)}(k)}/{dk}|_{k=k_F}$, the density of the one-electron state goes to zero near $k=k_F$. 

Figure\ref{fig:HFband} exhibits a large increase in bandwidth $\varepsilon_B = \varepsilon_{k_F} - \varepsilon_0$ in the HF band compared to the free-electron system. Considering that the bandwidth of the free electron 2D-UEL is equal to the Fermi energy $\varepsilon_F$, one can define the HF bandwidth as $\varepsilon_B = \varepsilon_F + \varepsilon^{(x)}_{k_F} - \varepsilon^{(x)}_0$. Since $\varepsilon^{(x)}_k$ is a negative function which monotonically increases by $k$ the HF energy of the system at the bottom of the band $k=0$ decreases by a larger amount than the energy of the system at $k=k_F$. Therefore, the HF prediction of $\varepsilon_B$ is larger than in the case of free electrons. As we show below, VMC and DMC reduce $\varepsilon_B$ by capturing the correlation energy.

\subsection{Comparing VMC and DMC \texorpdfstring{$m^*$}{mstar}}
\begin{figure}[htbp!]
    \centering
    \includegraphics[width=1.0\linewidth]{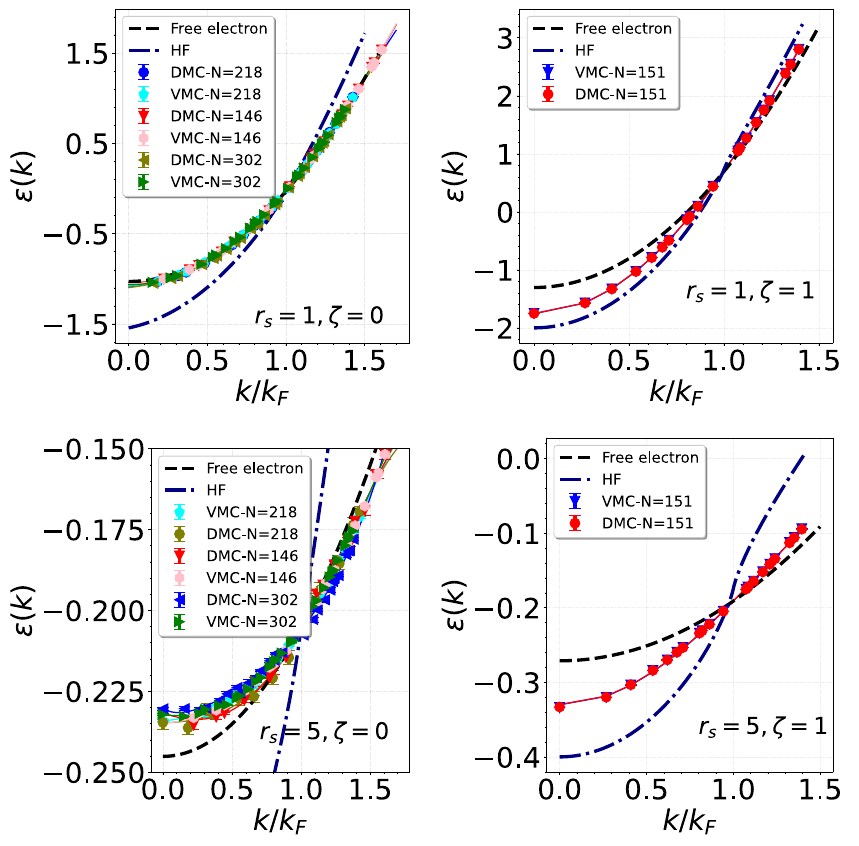}\\
    \caption{DMC and VMC energy bands (in a.u.)\ for paramagnetic and ferromagnetic 2D-UEL at $r_s=1$ and 5 obtained using system sizes $N=146$, 218, and 302 and $N=151$ for $\zeta=0$ and $\zeta=1$, respectively, and SJB wave function. A Pad\'{e} function is fitted to the VMC and DMC data. The free-electron and HF bands are offset to coincide with the fitted VMC bands at $k=k_F$.}
    \label{fig:para_band}
\end{figure}

The VMC and DMC energy bands of N-electron paramagnetic and ferromagnetic 2D-UEL with density parameters $r_s=1$ and 5 are shown in Fig.~\ref{fig:para_band}. The energy bands for paramagnetic 2D-UEL with density parameters $r_s=2,3,4$ are presented in the Supplementary Materials \cite{Suppl}. We also show the free-electron and HF energy bands in the thermodynamic limit in Fig.~\ref{fig:para_band}. The free-electron band is more accurate than the HF band, especially in lower density. The values of $m^*$ obtained using the VMC and DMC methods in exch $N$ are tabulated in the Supplementary Materials \cite{Suppl}. Our results indicate that $m^*$ obtained by the VMC and DMC methods agree with the error bars. All the main factors that can affect the results of the VMC and DMC for $m^*$ such as the nodal topology of the trial wave function, the optimization of the wave function, the fitting function and the DMC time step are discussed in the following sections. 

\subsection{Effect of nodal topology on \texorpdfstring{$m^*$}{mstar}}
In SJ wave function with bare Slater determinant defined in Eq.~(\ref{eq:slater}) the nodes are fixed and determined solely by the plane waves. BF modifies the particle coordinates to $X_i = r_i + \xi_i({r_j})$, where $\xi_i$ is the BF displacement, which is a function of the relative electron coordinates. So, in the SJB wave function the determinant becomes det$|\phi_i(X_j)| =$ det$ |\phi_i(r_j+\xi_j)|$ that effectively makes the nodal surface depend on the interelectronic correlation, not just the mean-field single-particle orbitals. In fact, the BF transformation includes an electron-electron correlation directly into the Slater determinant, which is absent in the HF Slater determinant. The FN approximation restricts the projected WF to have the same nodal surface as the trial WF, and therefore the quality of the nodal topology is crucial in the FN-DMC\@. Errors in the nodal surface lead to a variational bias. These errors significantly affect the QP energy band of 2D-UEL and can be substantially minimized by using the BF which is vital, especially at large $r_s$. 

\begin{figure}[!htb]
    \centering
    \includegraphics[width=1.0\linewidth]{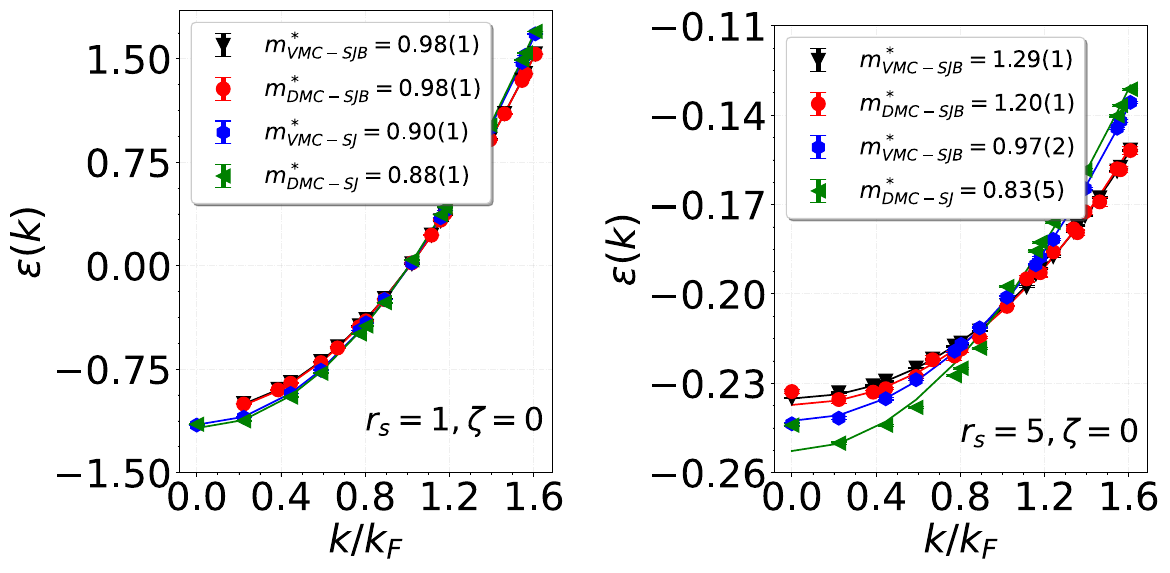}\\
    \caption{VMC and DMC energy bands of N-electron (N=146) paramagnetic 2D-UEL obtained by SJ and SJB wave functions with density parameters $r_s=1,5$. The effective mass $m^*$ is obtained using a quartic function fit.}
    \label{fig:SJ_SJB}
\end{figure}

We compared the VMC and DMC energy bands of paramagnetic 2D-UEL obtained using the SJ and SJB wave functions for the system size N = 146 and the density parameters $r_s=1$ and 5 (Fig.~\ref{fig:SJ_SJB}). Our results suggest that including the correlation within the Slater determinant using the BF transformation plays a crucial role in the effective mass of 2D-UEL\@. The pathological behavior in the VMC and DMC energy bands, which is caused by the divergence of the derivative of the HF band at $k=k_F$, dramatically affects the calculation of $m^*$. Decreasing this effect depends on the correlation energy retrieved in the QMC calculations, which can be achieved by using the BF transformation. Comparison of $m^*$ for $r_s=1,5$ obtained using the SJ and SJB wave functions shows that the correlation energy is more important at low density. The BF substantially improves the ground-state energy for both ground- and excited-state states. This improvement may cancel out when the QP energies are calculated. However, BF also changes the topology of the nodal surface of the wave function, which is vital to obtain the correct $m^*$. This can be observed by comparing the $m^*$ obtained by the VMC and DMC methods using the SJ and SJB wave functions. 

The SJ-VMC and DMC results for $m^*$ of 2D-UEL with $r_s=5$ are smaller than 1, suggesting that $m^*$ of paramagnetic 2D-UEL decreases with the reduction in density. However, the static correlation included in the Slater determinate through BF changes the results and predicts that $m^*$ of 2D-UEL with $r_s=5$ obtained by SJB-VMC and DMC are greater than one, predicting that $m^*$ of paramagnetic 2D-UEL increases when density decreases. Static correlation can be improved by using the full configuration interaction (FCI) method \cite{Shepherd}. Then one would expect that the $m^*$ of paramagnetic 2D-UEL at low density calculated by FCI would be larger than the SJB-VMC and DMC results. 

\subsection{Effect of wave function optimization on \texorpdfstring{$m^*$}{mstar}}
\begin{figure}[!htb]
    \centering
    \includegraphics[width=1.0\linewidth]{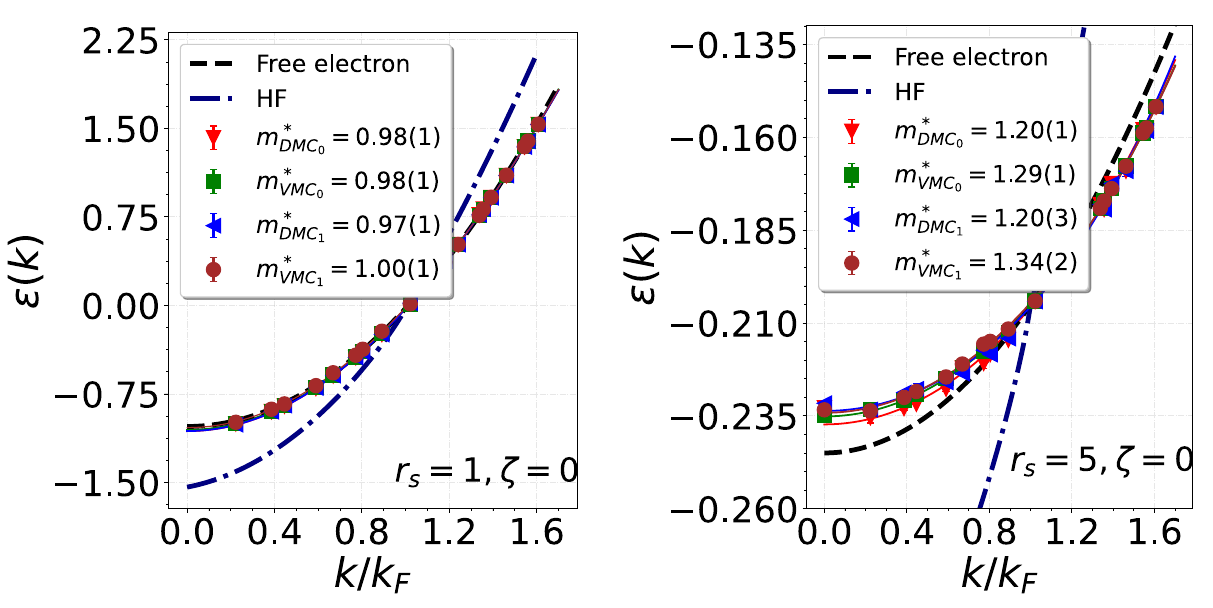} \\
    \caption{VMC and DMC energy bands of paramagnetic 2D-UEL obtained using the same WF for all $k$ vectors ($0$ index) and the optimized WF at each $k$ vector ($1$ index). The energy bands are calculated for $r_s=1,5$ with $N=146$ electrons in the simulation cell and SJB wave function. Quartic function fitting is used to calculate the effective mass $m^*$. The free-electron and HF bands are offset to coincide with the fitted DMC bands at $k=k_F$.}
    \label{fig:mStarOptWF}
\end{figure}

We compared the VMC and DMC energy bands of paramagnetic 2D-UEL obtained using two approaches. In the first approach, we optimized the real wave function for the system size $N=146$ and used it for all $k$-vectors in occupied and unoccupied energy bands. In the second approach, we used energy minimization to reoptimize the wave function in each $k$-vector. We calculated $m^*$ using the Pad\'{e} function fitting and compared the results for $r_s=1,2,5$. Figure~\ref{fig:mStarOptWF} shows the energy bands that are obtained using two approaches for $r_s=1,5$. The same results for $r_s=2$ are presented in the supplementary materials \cite{Suppl}. We found that $m^*$ obtained using VMC is more dependent on wave function optimization than DMC\@. The effect of wave function optimization in each $k$-vector on DMC effective mass is negligible. However, the energy minimization in each $k$-vector slightly increases the VMC effective mass. 

\subsection{Effect of fitting function on \texorpdfstring{$m^*$}{mstar}}
We carried out a least-square fit of a quartic function to VMC and DMC bands and compared the results with the ones which were obtained using Pad\'{e} function. The energy bands and effective masses obtained by fitting the quartic function are displayed in the supplementary materials \cite{Suppl}. We found that the quartic function provides a better fit than Pad\'{e} function. The fitting of the quartic function produces a slightly larger $m^*$ for all densities and system sizes. 

\subsection{Effect of DMC time step on \texorpdfstring{$m^*$}{mstar}}
The DMC time step $d\tau$ controls the imaginary-time propagation of walkers. Although the smaller $d\tau$ reduces the Trotter error (systematic errors due to discretizations), it requires more steps and consequently a longer simulation time to converge. Walkers represent configurations in an ensemble, and increasing them improves statistical averaging and reduces population control bias. When $d\tau$ is small, the walkers drift only slightly in each step and therefore the variance of the walker population increases slowly and the population control error is smaller. However, to keep the statistical fluctuations low, especially when branching is more frequent in small $d\tau$, a larger number of walkers $N_w$ is needed. The population control error scales roughly as $\sim (N_wd\tau)^{-1}$, meaning that decreasing $d\tau$ implies increasing $N_w$ to keep the bias small. In practice, the ground-state DMC energy of a system is obtained using the extrapolation of the DMC energy as a function of $d\tau$ to $d\tau \rightarrow0$. This procedure becomes very expensive for QP band simulations. Hence, it is important to systematically study the effect of $d\tau$ on the QP energy band and $m^*$.  

\begin{figure}[!htb]
    \centering
    \includegraphics[width=0.85\linewidth]{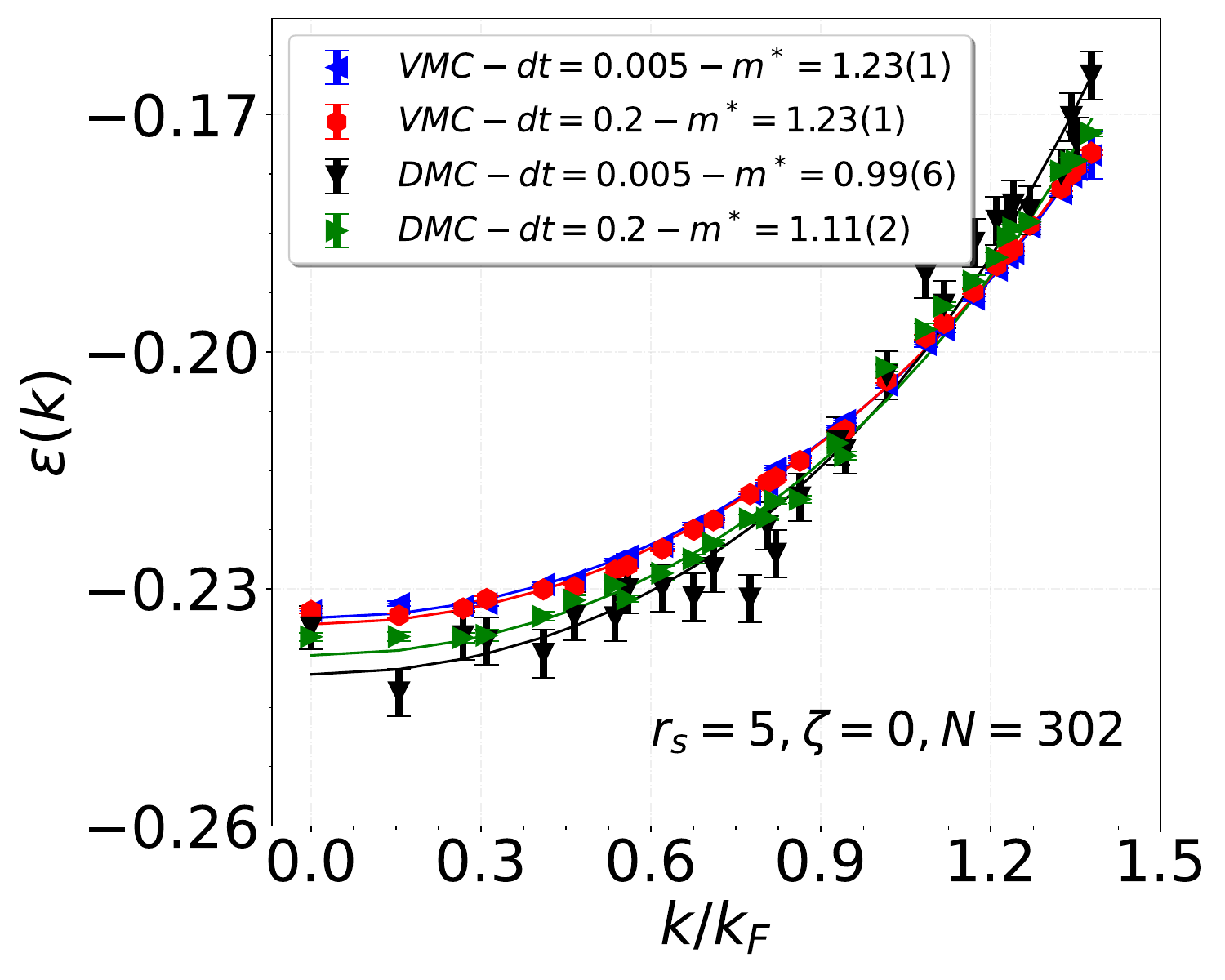}
    \caption{VMC and DMC energy bands of $N$-electron paramagnetic 2D-UEL ($N=302$) with density parameter $r_s=5$. The DMC energies are calculated using two time steps of 0.005 and 0.2 a.u. The effective mass is obtained using a quartic fitting.}
    \label{fig:DMCdt}
\end{figure}

We found that the DMC time step $d\tau$ has a crucial effect on $m^*$. We compared the DMC energy bands for system size $N=302$ and $r_s=5$ which are obtained with two different DMC time steps of 0.005 and 0.2 a.u.\ and the same number of walkers $N_w=2560$ and $25\times 10^3$ number of steps.  Figure~\ref{fig:DMCdt} displays the VMC and DMC energy bands of paramagnetic 2D-UEL with density parameter $r_s=5$ and $N=302$ electrons in the simulation cell. The fitting error for the DMC band obtained with $d\tau=0.005$ a.u.\ is much larger than that obtained with $d\tau=0.2$ a.u., resulting in a smaller $m^*$ with a large error bar. Our results show that the DMC time-step error is not canceled when the excitation energy is calculated for each vector $k$. 

\subsection{\texorpdfstring{$m^*$}{mstar} of 2D-UEL as a function of density}
\begin{figure}[htbp!]
    \centering
    \includegraphics[width=0.85\linewidth]{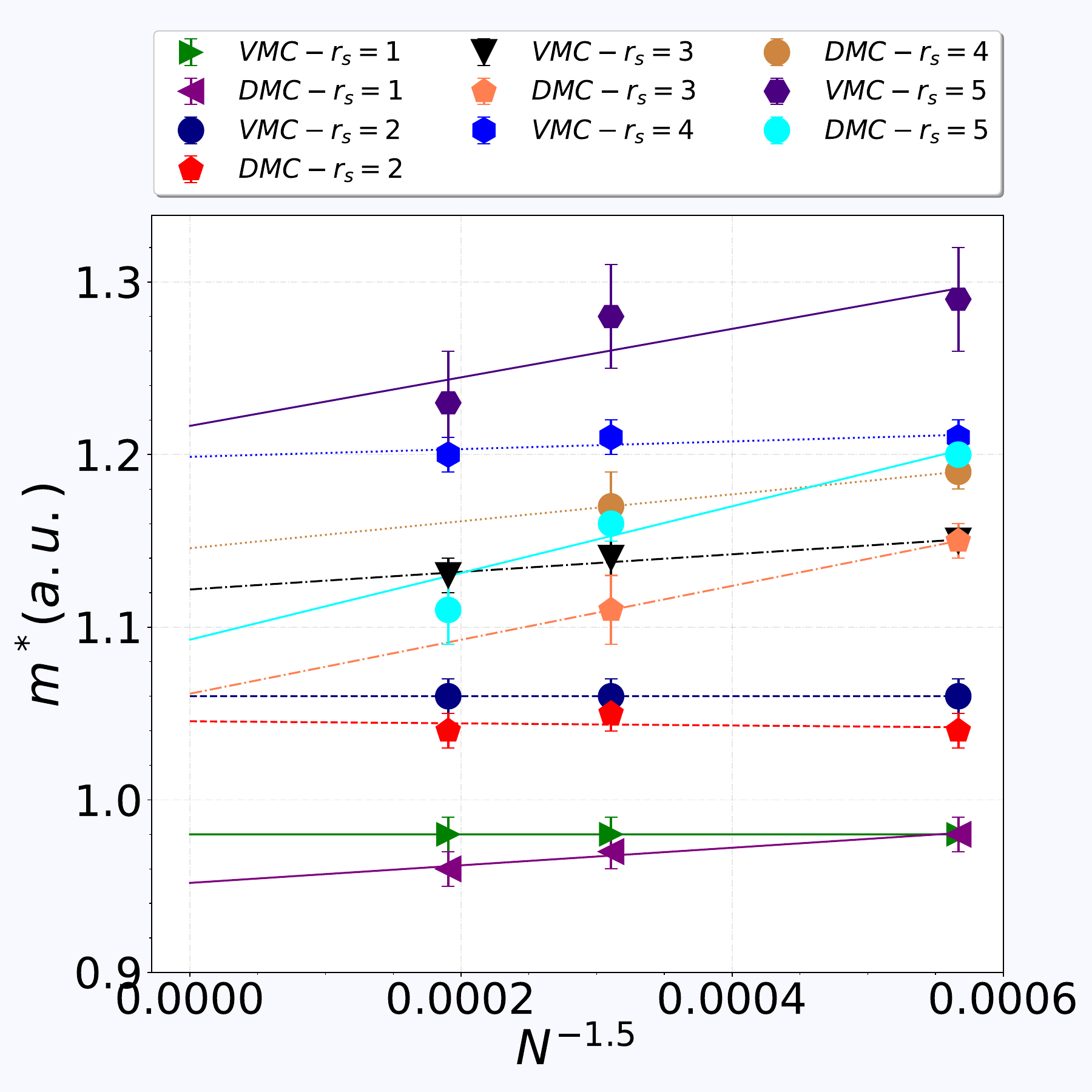}
    \caption{Quasiparticle effective mass $m^*$ of paramagnetic 2D-UEL obtained by VMC and DMC as a function of $N^{-1.5}$, where $N$ is the number of electrons in the simulation cell.}
    \label{fig:mstarvsN}
\end{figure}

The $m^*$ of the paramagnetic 2D-UEL obtained by the VMC and DMC methods is plotted against the system size, the number of electrons in the simulation cell $N$, in Fig.~\ref{fig:mstarvsN}. The systematic behavior of $m*$ as a function of $N^{-3/2}$ suggests that we can extrapolate $m^*$ to the infinite system size limit to reduce FS errors. We found that the FS errors in the VMC results for $m^*$ are less than DMC\@. In addition, FS errors increase as density is reduced. 

\begin{figure}[htbp!]
    \centering
    \includegraphics[width=0.85\linewidth]{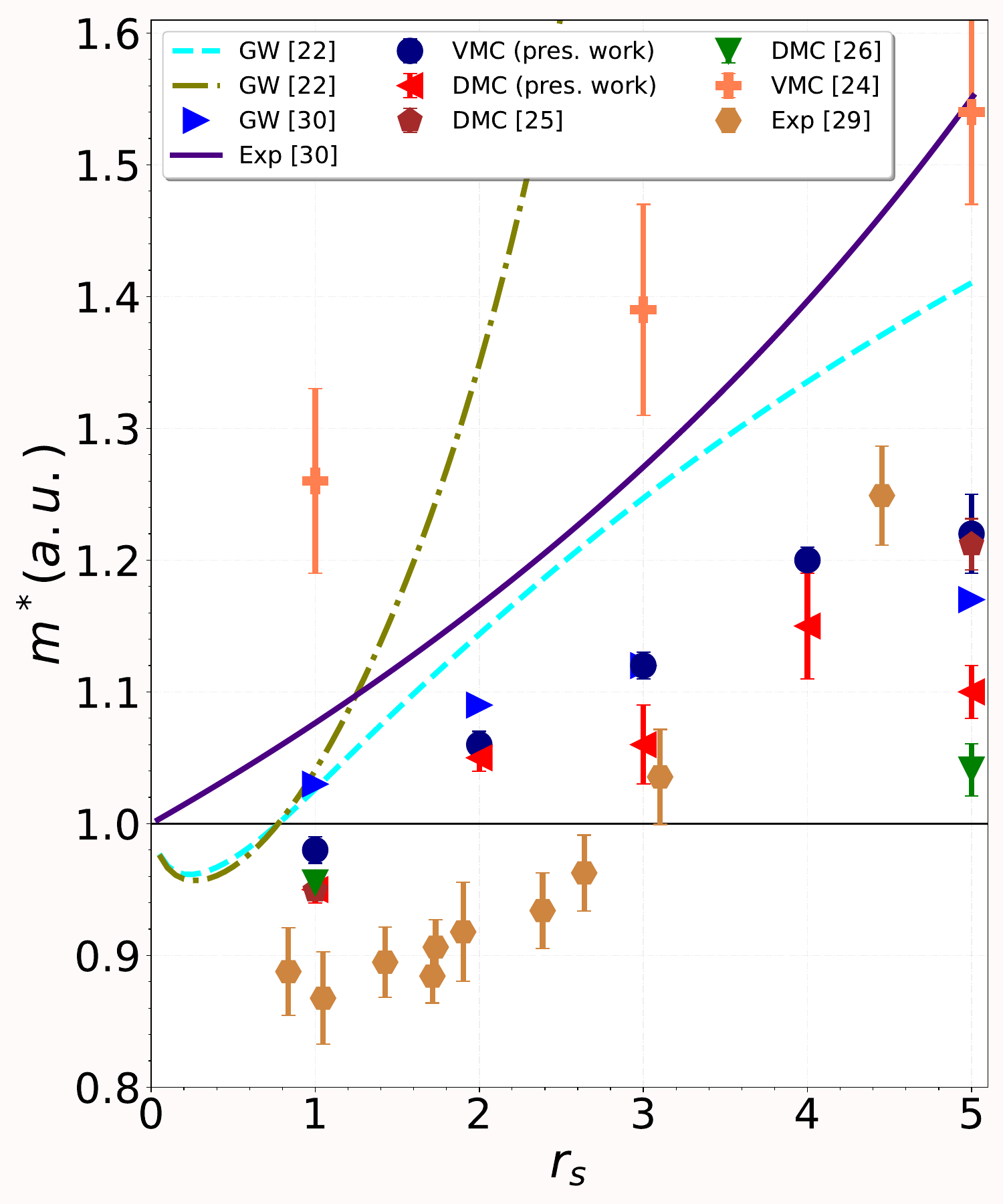}
    \caption{Quasiparticle effective mass $m^*$ as a function of density parameter $r_s$ for paramagnetic 2D-UEL\@. The $GW$ results were obtained using different approximations as discussed in the text.}
    \label{fig:mstarvsrspara}
\end{figure}

The quasiparticle effective mass $m^*$ against density parameter $r_s$ for paramagnetic 2D-UEL, which is obtained by experiments and theoretical methods, is shown in Fig.~\ref{fig:mstarvsrspara}.  $GW$ results were calculated using the RPA effective interaction \cite{Giuliani05}, and the Kukkonen-Overhauser effective interaction by solving the Dyson equation self-consistently or within the on-shell approximation \cite{Asgari06}. We compare our VMC and DMC results (pres.\ work) with previous VMC and DMC results \cite{Kwon93,Holzman09,Neilprb13}, and experiments \cite{Tan05,Smith72}. $GW$ calculations \cite{Asgari06,Giuliani05} show a steep increase in $m^*$ of paramagnetic 2D-UEL by decreasing the density, which is consistent with the initial experiment \cite{Smith72}. However, the $GW$ results can strongly be affected by the choice of effective interaction and whether or not the calculations are carried out self-consistently. 

Our VMC and DMC results for $m^*$ at high density $r_s=1$ suggest that the effective mass of paramagnetic 2D-UEL is close to 1, which means that the renormalization of the electron mass introduced by the electron-electron interaction is negligible. The VMC and DMC values of $m^*$ of the paramagnetic 2D-UEL in $r_s=1$ are 0.98(1) and 0.95(1) a.u., respectively. We have previously found that the DMC value of $m^*$ of the paramagnetic 3D-UEL in $r_s=1$ is 0.921(1) a.u.\ \cite{Azadi21}. It can then be concluded that the electron-electron interaction in 3D-UEL and its effect on the quasiparticle effective mass renormalization are stronger than in 2D-UEL\@. In addition, the $m^*$ of the paramagnetic 3D-UEL decreases as the density decreases \cite{Azadi21}, while the $m^*$ of the 2D-UEL increases when the density is reduced. 

\begin{figure}[htbp!]
    \centering
    \includegraphics[width=0.85\linewidth]{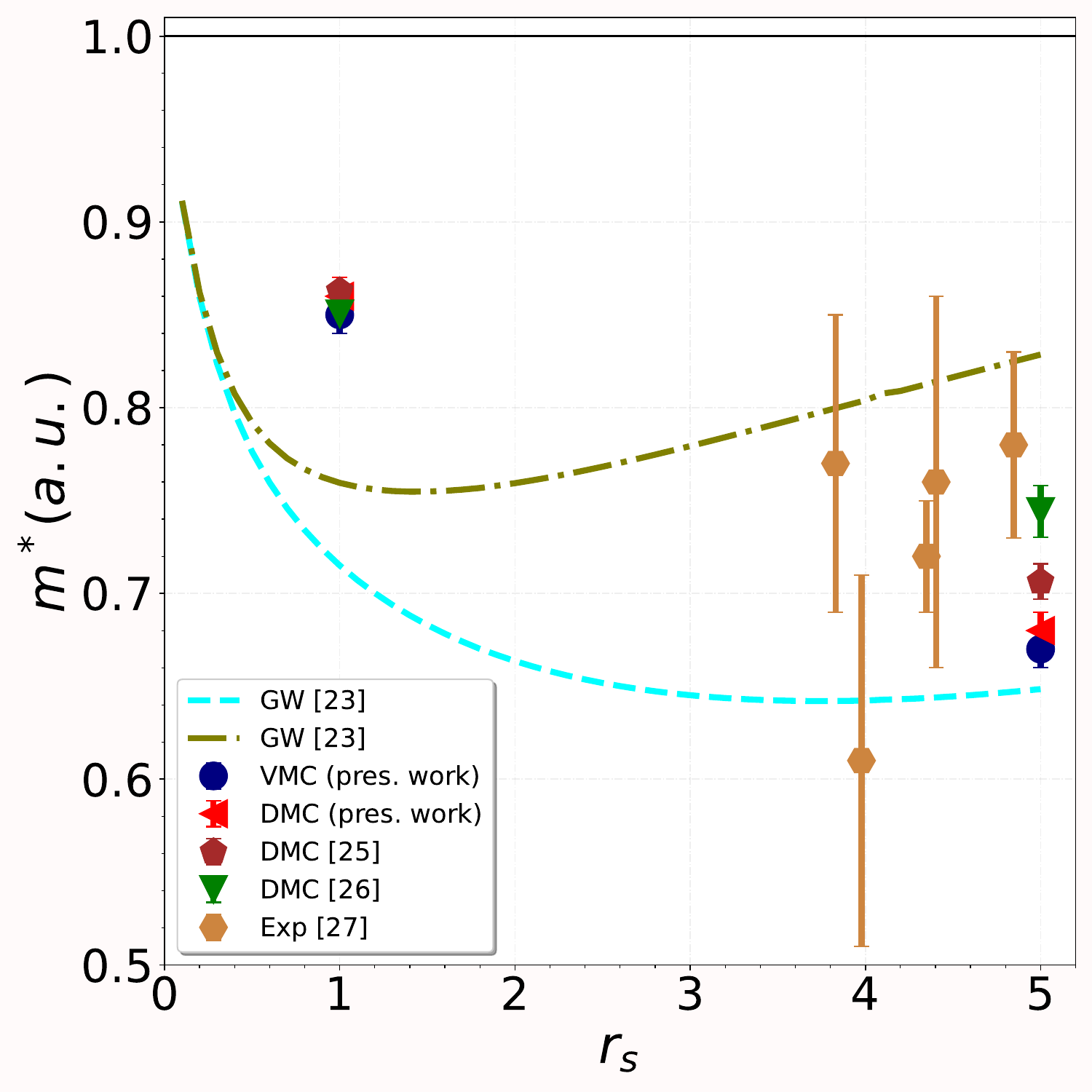}
    \caption{Quasiparticle effective mass $m^*$ as a function of density parameter $r_s$ for ferromagnetic 2D-UEL\@. The $GW$ results were obtained using different approximations as discussed in the text.}
    \label{fig:mstarvsrsfero}
\end{figure}

Figure~\ref{fig:mstarvsrsfero} shows the $m^*$ of ferromagnetic 2D-UEL as a function of density parameter $r_s$ obtained by the VMC, DMC, and $GW$ methods and compared to experiment. Our VMC and DMC data agree relatively well with the experimental data reported in Ref.~\onlinecite{Padmanabhan08}. It should be clarified that the 2D electron-liquid systems studied in the experiments are different from the ideal 2D UEL considered in this work. This is because the experimental samples have finite widths and are subject to disorder. Therefore, a precise quantitative comparison of our numerical data with available experimental data would be inappropriate.  Nevertheless, a qualitative behavior can provide useful information. Our QMC data and $GW$ results suggest a suppression of $m^*$ of ferromagnetic 2D-UEL within the range of density parameters of $1\leq r_s \leq5$. However, the difference between the $GW$ results obtained self-consistently and in the on-shell approximation is significant \cite{Asgari09}. The behavior of $m^*$ of ferromagnetic 2D-UEL as a function of density is similar to that of 3D-UEL \cite{Azadi21}, as in both cases $m^*$ decreases with the reduction in density. 

We discuss our QMC results for $m^*$ of 2D- (persent work) and 3D-UEL \cite{Azadi21}. From a perturbation theory point of view, effective mass renormalization comes from the real part of self-energy $\Sigma$. The parameters that can affect $m^*$ include exchange effects (dominate in high density, small $r_s$), correlation effects (increasingly important at low density, large $r_s$), dimensionality, polarization and screening. In paramagnetic systems, the Coulomb interaction is screened more efficiently in 3D, while the screening is weaker and less effective in 2D, especially at low density (large $r_s$). This screening in 3D-UEL improves at lower densities causing the reduction of $m^*$, while the correlation effects become stronger in 2D-UEL by increasing $r_s$ enhancing $m^*$. The evidence is the difference between $m^*$ of 2D- and 3D-UEL which are obtained using SJ and SJB wave functions, where the correlation energy is significantly improved by including the BF. This difference is negligible for 3D-UEL \cite{Azadi21}, while it is huge for 2D-UEL as can be observed in Fig.~\ref{fig:SJ_SJB}. Therefore, we argue that the competition between correlation and screening is the main reason for the difference between the behavior $m^*$ of 2D and 3D paramagnetic UEL as a function of $r_s$ at low densities. 

Another evidence is spin-polarization effects that suppress correlation in both 2D and 3D UEL, causing a reduction of $m^*$ as density decreases. The reason that correlation is suppressed in the spin-polarized case in comparison with the paramagnetic system is Pauli exclusion which prevents electrons from coming close. Hence, we can summarize that the electron-electron correlation enhances $m^*$ while the screening and spin-polarization decrease it.

\section{Conclusions}
We used the VMC and DMC methods to calculate $m^*$ of paramagnetic and ferromagnetic 2D-UEL within the metallic density range $1\leq r_s \leq 5$. We found that the VMC and DMC effective masses agree within the error bar. The $m^*$ of the paramagnetic 2D-UEL at the thermodynamic limit increases when the density decreases, opposite to the ferromagnetic system where $m^*$ suppresses by reducing density. We systematically investigated all the parameters that may affect the VMC and DMC results. We found that choosing a correct DMC time step is crucial in DMC calculation of the quasiparticle energy band of 2D-UEL\@. Optimization of the wave function in each wave vector $k$ has a negligible effect on the DMC results, but slightly increases  $m^*$ obtained with VMC\@. The $m^*$ of the paramagnetic 2D-UEL at high density $r_s=1$ and $N=146$ electrons in the simulation cell which is obtained using VMC and the optimized wave function in each wave vector $k$ is equal to the free electron $m^*$. Comparison of the results of VMC and DMC obtained by the SJ and SJB wave function indicates that the correlation between electrons plays a crucial role in the behavior of $m^*$ as a function of density. The $m^*$ of the paramagnetic 2D-UEL obtained by the SJ wave function is smaller than one, while including the electron correlation in the Slater determinant by using the BF transformation gives a larger value than one for $m^*$. Comparison of all VMC and DMC results suggests that VMC is a reliable alternative to DMC for studying excite states as long as the trial wave function is well optimized. Especially in the case of large system sizes where the cost of DMC simulation is much more than VMC\@. 

\begin{acknowledgments}
S.\ A., A.\ P., and M.\ S.\ B.\ acknowledge the support of the Leverhulme Trust under the grant agreement RPG-2023-253. S.\ A., M.\ S.\ B.\, and R.\ V.\ B.\ gratefully acknowledge the Research Infrastructures at the Center for Computational Materials Science at the Institute for Materials Research for allocations on the MASAMUNE-IMR supercomputer system (Project No. 202112-SCKXX-0510). M.\ S.\ B.\ and R.\ V.\ B.\ are grateful to E-IMR center at the Institute for Materials Research, Tohoku University, for continuous support.

\end{acknowledgments}

\bibliography{main}

\end{document}